\documentclass[aps,prd,twocolumn,nofootinbib,superscriptaddress,groupedaddress,showkeys]{revtex4}  % for review and submission
\usepackage{graphicx}  % needed for figures
\usepackage{dcolumn}   % needed for some tables
\usepackage{bm}        % for math
\usepackage{amssymb}   % for math
\usepackage{amsmath}
\usepackage{xcolor}
\usepackage{bbold}
\usepackage{amsthm}
\newtheorem{theorem}{Theorem}
\newcommand{\be}{\begin{equation}}\newcommand{\ee}{\end{equation}}
\newcommand{\bea}{\begin{eqnarray}}\newcommand{\eea}{\end{eqnarray}}
\newcommand{\brr}{\begin{array}}\newcommand{\err}{\end{array}}
\newcommand{\bit}{\begin{itemize}}\newcommand{\eit}{\end{itemize}}
\newcommand{\ben}{\begin{enumerate}}\newcommand{\een}{\end{enumerate}}

\newcommand{\bbm}{\begin{bmatrix}}\newcommand{\ebm}{\end{bmatrix}}
\newcommand{\ba}{\begin{array}}
\newcommand{\ea}{\end{array}}

\newtheorem{mydef}{Definition}
\newtheorem{Lemma}{Lemma}
\newcommand{\bd}{\begin{mydef}} \newcommand{\ed}{\end{mydef}}
\newcommand{\bthe}{\begin{theorem}} \newcommand{\ethe}{\end{theorem}}
\newcommand{\ble}{\begin{Lemma}} \newcommand{\ele}{\end{Lemma}}

\definecolor{darkred}{rgb}{.8,0,0}

\definecolor{darkblue}{rgb}{0,0,.7}

\def\ph{\varphi}
\def\lan{\langle}
\def\lf{\left}

\def\non{\nonumber}\def\ran{\rangle}

\def\ri{\right}
\def\al{\alpha}\def\bt{\beta}\def\ga{\gamma}
\def\De{\Delta}

\def\la{\lambda}

\def\1{{_{1}}}\def\2{{_{2}}}

\def\noHe0{:\;\!\!\;\!\!:H_e(0):\;\!\!\;\!\!:}
\def\noHm0{:\;\!\!\;\!\!:H_\mu(0):\;\!\!\;\!\!:}

\def\lan{\langle}
\def\lf{\left}

\def\non{\nonumber}
\def\ran{\rangle}

\def\ri{\right}

\def\al{\alpha}\def\bt{\beta}\def\ga{\gamma}
\def\De{\Delta}

\def\la{\lambda}

\def\1{{_{1}}}\def\2{{_{2}}}

\begin{document}

%%%%%%%%%%%%%%%%%%%%%%%%%%%%%%%%%%%%%%%%%%%%%%%%%%%%%%%%%%%%%%%%%%%%%%%%%%%%%%%%%%%%%%%%%%%%%%%%%%%%%%%%%%%%%%%%%%%
\title{Neutrino oscillations in extended theories of gravity}
%%%%%%%%%%%%%%%%%%%%%%%%%%%%%%%%%%%%%%%%%%%%%%%%%%%%%%%%%%%%%%%%%%%%%%%%%%%%%%%%%%%%%%%%%%%%%%%%%%%%%%%%%%%%%%%%%%%
\author{Luca~Buoninfante}
\email{lbuoninfante@sa.infn.it}
\affiliation{Dipartimento di Fisica, Universit\`a di Salerno, Via Giovanni Paolo II, 132 84084 Fisciano, Italy}
\affiliation{INFN Sezione di Napoli, Gruppo collegato di Salerno, Italy}
\affiliation{Van Swinderen Institute, University of Groningen, 9747 AG, Groningen, The Netherlands}
\author{Giuseppe Gaetano Luciano}
\email{gluciano@sa.infn.it}

\affiliation{INFN Sezione di Napoli, Gruppo collegato di Salerno, Italy}
\author{Luciano Petruzziello}
\email{lpetruzziello@na.infn.it}

\affiliation{Dipartimento di Fisica, Universit\`a di Salerno, Via Giovanni Paolo II, 132 84084 Fisciano, Italy}
\affiliation{INFN Sezione di Napoli, Gruppo collegato di Salerno, Italy}
\author{Luca~Smaldone}
\email{lsmaldone@sa.infn.it}
\affiliation{INFN Sezione di Napoli, Gruppo collegato di Salerno, Italy}
%
%\date{\today}
\vspace{3mm}

\begin{abstract}
We study neutrino oscillations 
within the framework of extended theories 
of gravity. Based on the 
covariant reformulation of Pontecorvo's formalism, 
we evaluate the oscillation 
probability of neutrinos propagating  
in static spacetimes described 
by gravitational actions quadratic in the 
curvature invariants. Calculations are
carried out in the two-flavor approximation, 
for oscillations both in vacuum and matter. 
It is shown that the neutrino phase is sensitive 
to the violation of the strong equivalence principle.  
By way of illustration, we specialize our analysis  
to various extended models of gravity in order both
to quantify such a violation and to understand
how the characteristic free parameters of these models
affect the neutrino phase.
The possibility to fix new bounds on these parameters and 
to constrain extended theories of gravity 
is finally discussed.
\end{abstract}


\maketitle

%%%%%%%%%%%%%%%%%%%%%%%%%%%%%%%%%%%%%%%%%%%%%%%%
\section{Introduction}
Neutrinos are among the most enigmatic entities  
in particle physics. Because of their zero 
charge and extremely small masses, 
they impinge on matter (almost) only by 
the weak interaction. Such an elusive nature 
justifies the nearly three-decades delay 
between Pauli's prediction of the existence of (anti-)neutrino in 1930 
and its real detection by Reines and Cowan Jr \cite{ReiCow}
in 1956.
Since then, neutrino physics has been largely addressed,
drawing even more attention
after Pontecorvo's pioneering idea of 
flavor mixing and oscillations~\cite{Pontecorvo}.
Although a firm treatment of 
these phenomena has now been set up
both at theoretical~\cite{giu} and experimental~\cite{exp} levels, 
in vacuum~\cite{Kayser} and in matter~\cite{matter}, 
such puzzling questions as the correct quantum field theoretical definition of flavor states~\cite{NeutPheno,bla}, the nature of neutrino masses (Dirac or Majorana) \cite{DirMaj} and the dynamical origin of the non-vanishing neutrino masses and mixings~\cite{Blasone:2018hah, Blasone2} are  
still under investigation.

All of the above is set in flat spacetime.
Neutrino oscillations in the presence of gravity 
were first studied by Stodolsky~\cite{Stodo}, and their
relevance in cosmology and astrophysics was
later pointed out in Refs.~\cite{relevance,Cardall}. 
Recently, a similar analysis in accelerated frames 
has burst into the spotlight~\cite{acc,Pitjiz,epl} 
in connection with the controversy on the asymptotic nature
of mixed neutrinos in the decay of accelerated protons~\cite{controversy}.
These studies, however, have been carried out within the
framework of Einstein's General Relativity  (GR).
Despite providing the most successful description of 
gravitational interaction~\cite{will},
it is nowadays commonly thought 
that GR might not be the ultimate theory,  
because of its incompleteness at short distances or, in other words, at high energies (think of classical singularities and lack of renormalizability), and 
its failure to explain 
such issues as the cosmic inflation or the possible
existence of dark matter and dark energy.
This paves the way
for a strenuous search of new models~\cite{capozdela} 
that may encompass these problems in 
a self-consistent scheme, preserving at the same time
the positive results of GR.

Among all the extended theories of gravity formulated in the years, 
the most straightforward approaches are the so-called
\emph{quadratic theories}, which consist of generalizing the
Einstein-Hilbert gravitational action by including contributions 
quadratic in the curvature invariants. In this context, worthy of note 
are the results achieved by Stelle~\cite{Stelle}, 
who realized that a description of gravity arising from the 
Einstein-Hilbert action containing the squared scalar curvature and 
squared Ricci tensor, 
$\mathcal R^{2}$ and $\mathcal{R}_{\mu\nu}\mathcal{R}^{\mu\nu}$, is 
power-counting renormalizable. 
However, such a theory lacks of
predictability above a certain 
cutoff, which is given by the
mass of a spin-$2$ ghost degree of freedom appearing in the
theory when the standard quantization is adopted.
Developments have been subsequently 
gained in the Starobinsky model of cosmic inflation~\cite{Starob}, which only 
involves the $\mathcal{R}^2$ term, and also in other 
scenarios~\cite{Capozziello:2014mea}.
Interesting results have been also highlighted 
in \emph{non-local} quadratic theories~\cite{Biswas:2005qr,Modesto:2011kw,Biswas:2011ar,Biswas:2013cha,Biswas:2016etb,Koshelev:2017bxd,Buoninfante:2018mre,Buoninfante:2018xiw}.

Understanding which of the above
extended theories may be considered as
the best candidate to generalize GR and, as a consequence, how it 
affects physical phenomena is certainly a crucial task \cite{will,capozdela}.
For instance, a recent attempt to fulfill this aim has been made
in the context of Casimir effect in Refs.~\cite{Buoninf,petr}, 
where non-trivial bounds on the free parameters 
appearing in  such theories
have been inferred by the evaluation of Casimir energy density and 
pressure. 
In the present paper, we will face this issue 
by analyzing neutrino flavor oscillations both in matter and vacuum and computing
the correction to the quantum mechanical phase
arising from the extra terms in the gravitational action. 
In this regard, we remark
that a similar analysis has been carried out in Brans-Dicke theory in
Ref.~\cite{Capozziello:1999qm} and in other extended
models in Ref.~\cite{capozdela}.

On the other hand, differently from the previous approaches, we will also 
discuss the possibility to pinpoint phenomenological implications
of the strong equivalence principle (SEP) on neutrino propagation, as already
investigated in Refs.~\cite{strongequivprinc}.
Indeed, there is a common agreement on the SEP violation 
occurrence in some extended models of gravity~\cite{will}. For instance, in the context of 
the aforementioned Brans-Dicke theory, one can evaluate the inertial and gravitational
mass of the source of gravity and notice that the presence of the dynamical scalar field is the responsible for the discrepancy between the two terms~\cite{ohan,Petrozzolo}. Such a violation can be also extended to $f(\mathcal{R})$ models, in light of the close bond they share with  scalar-tensor theories, which has been the subject of an intense
line of research (i.e. see for example Refs.~\cite{capozdela,stfr}).
Motivated by these ideas, one of our primary aims is to seek the 
contribution to the neutrino oscillation phase that can be 
associated to SEP violation. 
%In addition to that, we will argue a possible origin of such a term starting from general consideration revolving around the neutrino Dirac Hamiltonian.

The layout of the paper is the following. In Section~\ref{cs}
we analyze the standard formalism of vacuum neutrino oscillations 
in Minkowski framework. We also introduce the covariant
formulation of Ref.~\cite{Cardall}. The same considerations
are then extended to the case of oscillations in matter. 
Section~\ref{qtg} is devoted
to a review of the most important features of some quadratic
theories of gravity. Corrections to the neutrino quantum mechanical
phase and to the related oscillation probability 
are explicitly calculated in Section~\ref{ope}.
%Moreover, the possibility of recognizing signatures of SEP violation in the Dirac Hamiltonian for neutrinos is investigated. 
Aside from this, we discuss the possibility to fix 
constraints on the free parameters
appearing in the considered theories.
Section~\ref{example} contains a thorough application of the aforementioned general notions
to several quadratic models of gravity whose relevance has been proved
in a quantum field theoretical framework. Moreover, we 
explicitly point out the contribution to the covariant oscillation
phase that is directly related to SEP.
Concluding remarks can be found in Section~\ref{conc}.

Throughout the work we assume natural units $\hbar=c=1$ 
and the mostly negative metric convention, $\eta_{\mu\nu}={\rm diag}[1-1,-1,-1]\,.$ Furthermore, 
we consider a simplified two-flavor model for neutrinos:
the obtained results can be easily extended to a more general
three-flavor description with $CP$ violation.
%%%%%%%%%%%%%%%%%%%%%%%%%%%%%%%%%%%%%%%%%%%%%%%%%%%%%%%%%%%%%%%%%%%%%%%%%%%
\section{Neutrino oscillations in curved spacetime}  
\label{cs}
%%%%%%%%%%%%%%%%%%%%%%%%%%%%%%%%%%%%%%%%%%%%%%%%%%%%%%%%%%%%%%%%%%%%
\subsection{Vacuum oscillations} \label{csA}
Let us consider a flavor neutrino emitted
via weak interaction at a generic spacetime point.
According to Pontecorvo's quantum mechanical 
formalism~\cite{Pontecorvo}, the flavor state
$|\nu_\alpha\rangle$ $(\alpha=e,\mu)$
can be expressed as a superposition of 
the mass eigenstates $|\nu_k\rangle$ $(k=1,2)$
as\footnote{In what follows, we denote
flavor (mass) indices by greek (latin) indexes.}
\be
\label{flavneutr}
|\nu_\al\ran=\sum_{k=1,2} U_{{\alpha k}}\hspace{0.1mm}(\theta) \, |\nu_k\ran\,,
\ee
where $U_{\al k}(\theta)$ is the
generic element of the Pontecorvo mixing matrix
\begin{equation}
\label{PMM}
U(\theta)=
\begin{pmatrix}
\cos\theta&\sin\theta\\
-\sin\theta&\cos\theta
\end{pmatrix}.
\end{equation}
The states that indeed propagate are
the mass ones, whose energy $E_k$
and three-momentum $\vec{p}_{k}$
are related by the usual mass-shell condition
$E^2_k=m^2_k+|\vec{p}_{k}|^2$.

In Minkowski 
spacetime, the propagation of the
state $|\nu_k\rangle$ from a point $A(t_A,\vec{x}_A)$
to a point $B(t_B,\vec{x}_B)$ can be described
by a plane wave as
\be
|\nu_k(x)\ran=\exp{[-i\varphi_k(x)]}\hspace{0.2mm}|\nu_k\rangle,
\ee
where the phase $\Phi_k$ is defined as
\be
\label{phase}
\varphi_k=E_k(t_B-t_A)-\vec{p}_k\cdot(\vec{x}_B-\vec{x}_A)\,.
\ee
Therefore, by using Eqs.~\eqref{flavneutr} and~\eqref{phase}, 
the probability that a neutrino produced with flavor $\alpha$
at the point $A$ is detected with flavor $\beta$ at the point $B$ 
takes the form
\begin{eqnarray}
\label{probo}
\non
\mathcal{P}_{\al\rightarrow \bt}&=&{\Big|\lan \nu_\bt(t_B,\vec{x}_B)|\nu_\al(t_A,\vec{x}_A)\ran\Big|}^2\\[2mm]
&=&\sin^2 (2 \theta)\, \sin^2 \lf(\frac{\varphi_{12}}{2}\ri),
\end{eqnarray}
where the phase-shift is given by 
$\varphi_{12}=\varphi_1-\varphi_2$. 

For relativistic neutrinos, 
by assuming the mass eigenstates to be energy eigenstates 
with a common energy $E$,
one can show that 
\be\label{minkp}
\varphi_{12}\simeq\frac{\Delta m^2}{2E%_*
}\hspace{0.2mm}L_p\,,
\ee
where $\Delta m^2\equiv |m^2_2-m_1^2|$ is the
mass-squared difference and
$L_p=|\vec{x}_B-\vec{x}_A|$ is the distance travelled by neutrinos.

The above formalism can be
generalized in a straightforward way 
to curved spacetime by rewriting the phase 
\eqref{phase} as the eigenvalue of the
covariant operator~\cite{Cardall}
\be
\label{phasep}
\Phi=\int_{\lambda_A}^{\lambda_B} P_{\mu}\frac{d x^\mu_{null}}{d \lambda}\hspace{0.2mm}d\lambda\,,
\ee
where $P_\mu$ is the generator of spacetime
translations of neutrino mass eigenstates and $dx^\mu_{null}/d\lambda$ 
is the null tangent vector to the neutrino worldline parameterized by $\lambda$.
For neutrino propagating in flat spacetime, the above relation
recovers Eq.~\eqref{phase}, as it should be.

The quantity $P_{\mu}\,d x^\mu_{null}/d\lambda$ in Eq.~\eqref{phasep}
can be calculated starting from 
the covariant Dirac equation for a doublet 
of spinors $\nu$ of different masses~\cite{Weinberg:1995mt}
\begin{equation}
\label{eqn:covdireq}
\Big[i\gamma^{\hat{a}}\hspace{0.2mm}e^{\mu}_{\hat{a}}\left(\partial_\mu\,+\,\Gamma_\mu\right)\,-\,M\Big]\nu\,=\,0\,,
\end{equation}
where $M=\mathrm{diag}[m_1,m_2]$  
and $\gamma^{\hat{a}}$ are the Dirac matrices.
The general curvilinear and locally inertial sets of coordinates 
are denoted without and with hat, respectively, and they are related by the 
vierbein field $e^{\mu}_{\hat {a}}$. The explicit expression for
the Fock-Kondratenko connection is $\Gamma_\mu=\frac{1}{8}\left[\gamma^{\hat {b}},\gamma^{\hat{c}}\right]\hspace{-0.5mm}e^{\nu}_{\hat{b}} \hspace{0.2mm}e_{\hat{c}\hspace{0.2mm}\nu;\mu}\,,$ 
where the semicolon stands for the covariant derivative.

Note that the Dirac equation~\eqref{eqn:covdireq}
can be simplified by means of the following relation~\cite{Cardall}:
\begin{equation}
\gamma^{\hat{a}}\hspace{0.1mm}e^{\mu}_{\hat{a}}\hspace{0.1mm}\Gamma_\mu\,=\,\gamma^{\hat{a}}\hspace{0.1mm}e^{\mu}_{\hat{a}}\left\{i\hspace{0.1mm}A_{G\mu}\left[-\,{g}^{-1/2}\hspace{0.2mm}\frac{\gamma^5}{2}\right]\right\}\,,
\end{equation}
where $g\equiv|\mathrm{det}\,g_{\mu\nu}|$, $\gamma^5=i\gamma^{\hat{0}}\gamma^{\hat{1}}\gamma^{\hat{2}}\gamma^{\hat{3}}$
and the vector potential $A^{\mu}$ is given by
\begin{equation}
\label{eqn:A}
A^{\mu}_G\,=\,\frac{1}{4}\hspace{0.3mm}{g}^{1/2}\hspace{0.3mm}e^{\mu}_{\hat{a}}\hspace{0.3mm}\epsilon^{{\hat{a}}{\hat{b}}{\hat{c}}{\hat{d}}}\left(e_{{\hat{b}}\nu,\sigma}\,-\,e_{{\hat{b}}\sigma,\nu}\right)e^{\nu}_{\hat{c}} e^{\sigma}_{\hat{d}}\,.
\end{equation}
Here $\epsilon^{{\hat{a}}{\hat{b}}{\hat{c}}{\hat{d}}}$ is the totally antisymmetric Levi-Civita symbol
with component $\epsilon^{{\hat{0}}{\hat{1}}{\hat{2}}{\hat{3}}}\,=\,+1$. 

In the above setting, the momentum operator 
$P_\mu$ used to calculate the neutrino oscillation phase 
can be derived from the generalized mass-shell relation
\begin{equation}
\label{eqn:genmascond}
\left(P^{\mu}\,+\,A^\mu_{G}P_{L}\right)\left(P_{\mu}\,+\,A_{G\mu}P_{L}\right)\,=\,M^2\,,
\end{equation}
where $P_L \equiv (1-\ga_5)/2$ is the left-hand projector and 
we have added a term proportional to the identity without any physical consequences~\cite{Cardall}. By neglecting terms of order $\mathcal{O}(A_G^2)$ and $\mathcal{O}{(A_GM^2)}$ and considering relativistic neutrinos, we then obtain
\begin{equation}
\label{eqn}
P_\mu\hspace{0.3mm}\frac{d x^\mu_{null}}{d \lambda}\,=\,\left(\frac{M^2}{2}\,-\,\frac{d x^\mu_{null}}{d \lambda}A_{G\mu}P_{L}\,\right),
\end{equation}
where we have required $P^i\approx p^i$ and $P^0=p^0$~\cite{Cardall}.
In this regard, we emphasize that 
$E\equiv P_0=g_{0\nu}P^{\nu}$.

Finally, by denoting the differential proper distance at constant $t$
by $d \ell$, we can write
\bea \non 
d\lambda  & = & d \ell{\left(\hspace{-1mm}-g_{ij}\frac{d x^i}{d \lambda}\frac{d x^j}{d \lambda}\right)}^{\hspace{-1mm}-\frac{1}{2}}\hspace{-2mm}\\[2mm] \label{line}
& = & d \ell{\left[g_{00}\lf(\frac{d x^0}{d \lambda}\ri)^2\hspace{-1.5mm}+2g_{0i}\frac{d x^0}{d \lambda}\frac{d x^i}{d\lambda}\right]}^{-\frac{1}{2}}\hspace{-3mm} ,
\eea
where we have exploited the condition of null trajectory $d s^2=0$.

%%%%%%%%%%%%%%%%%%%%%%%%%%%%%%%%%%%%%%%%%%%%%%%%%%%%%%%%%%%%%%%%%%%%%%%%%%%
\subsection{Matter effects}
\label{matef}
The weak-field approximation, which we use along this paper, is suitable for the description of solar and supernovae neutrinos~\cite{Capozziello:1999qm,relevance, Grossman}. However, in these cases, such matter effect as Mikheyev-Smirnov-Wolfenstein (MSW) effect \cite{matter} cannot be disregarded. Following the treatment of Ref. \cite{Cardall}, we shall treat these extra contributions in a similar fashion to the gravity-induced corrections computed above. 

Let us assume that only electron neutrinos weakly interact with an electron background fluid. In this case, the generalized mass-shell relation takes the form
\begin{equation}
\label{eqn:genmascondmat}
\left(P^{\mu}\,+\,A^\mu_{f} P_L\right)\left(P_{\mu}\,+\,A_{f\mu} P_L\right)\,=\,M^2_f \, , 
\end{equation}
where $M^2_{f}$ $\equiv$ $U(\theta)\hspace{0.2mm}M^2\hspace{0.2mm}U^\dag(\theta)$ and 
\be
A^\mu_{f} \,\equiv \, \begin{pmatrix} -\sqrt{2} \, G_F \, N^{\mu}_e & 0 \\ 0 & 0  \end{pmatrix}\,, 
\ee
is the interaction term in the flavor basis, 
$G_F$ is the Fermi constant and $N^{\mu}_e=n_e u^\mu$ 
is the number current of the electron fluid. 
Here, $n_e$ and $u^\mu$ are the electron 
density in the fluid rest frame and the fluid's four-velocity, respectively. 

Finally,  by taking into account both geometric and matter effects, 
the generalization of Eq.~\eqref{eqn} reads
\begin{equation}
\label{eqnmat}
P_\mu\hspace{0.3mm}\frac{d x^\mu_{null}}{d \lambda}\, = \, \left(\frac{M^2}{2}\, - \,\frac{d x^\mu_{null}}{d \lambda}A_{\mu} P_L\,\right). 
\end{equation}
%
%%%%%%%%%%%%%%%%%%%%%%%%%%%%%%%%%%%%%%%%%%%%%%%%%%%%
\section{Quadratic theories of gravity}
\label{qtg}

In this Section, we introduce a wide class of extended theories of gravity for which the neutrino oscillation phenomenon will be studied.

Let us consider the following gravitational action which is the most general parity-invariant and torsion-free action around maximally symmetric backgrounds \cite{Biswas:2005qr,Biswas:2016etb},
\begin{eqnarray}
\nonumber
S&\hspace{-1mm}=\hspace{-1mm}& \frac{1}{2\kappa^2}\int\left\lbrace \mathcal{R}+\frac{1}{2}\Big[\mathcal{R}\mathcal{F}_1(\Box)\mathcal{R}\,+\,\mathcal{R}_{\mu\nu}\mathcal{F}_2(\Box)\mathcal{R}^{\mu\nu}+\right.\\[2mm]
\hspace{-5mm}&&+\mathcal{R}_{\mu\nu\rho\sigma}\mathcal{F}_3(\Box)\mathcal{R}^{\mu\nu\rho\sigma}\Big]\bigg\rbrace\hspace{0.1mm}\sqrt{-g}\hspace{1mm}d^4x,
\label{quad-action}
\end{eqnarray}
where $\kappa\equiv \sqrt{8\pi G}=1/M_p$ is the inverse of the reduced Planck mass, $\Box=g^{\mu\nu}\nabla_{\mu}\nabla_{\nu}$ is the curved d'Alembertian and the three differential operators $\mathcal{F}_i(\Box)$ are generic functions of $\Box:$
\begin{equation}
\mathcal{F}_i(\Box)=\sum\limits_{n=0}^{N}f_{i,n}\Box^n,\,\,\,\,\,\,\,\,\,i=1,2,3.
\end{equation}
Here, we deal with both positive ($n>0$) and negative ($n<0$) powers of the d'Alembertian, namely we analyze both ultraviolet and infrared modifications of Einstein's GR. When $N$ is finite ($N<\infty$) and $n>0$, we have a local theory of gravity whose derivative order is $2N+4$, while if $N=\infty$ and/or $n<0$ the corresponding gravitational theory is nonlocal and the form-factors $\mathcal{F}_i(\Box)$ are non-polynomial differential operators of $\Box.$

Since we are interested in computing and studying the neutrino oscillation phase in presence of a weak gravitational field, we can work in the linear regime by expanding the action in Eq.~\eqref{quad-action} around the Minkowski background
\begin{equation}
g_{\mu\nu}=\eta_{\mu\nu}+\kappa h_{\mu\nu}\,,\label{lin-metric}
\end{equation}
where $h_{\mu\nu}$ is the metric perturbation. 

In our perturbative approach, we truncate the action in Eq.~\eqref{quad-action} at order\footnote{In this regime, the term $\mathcal{R}_{\mu\nu\rho\sigma}\mathcal{F}_3(\Box)\mathcal{R}^{\mu\nu\rho\sigma}$ in Eq.~\eqref{quad-action} can be neglected. Indeed, the following identity holds true:
$$
\mathcal{R}_{\mu\nu\rho\sigma}\Box^n\mathcal{R}^{\mu\nu\rho\sigma}=4\mathcal{R}_{\mu\nu}\Box^n\mathcal{R}^{\mu\nu}-\mathcal{R}\Box^n\mathcal{R}+\mathcal{O}(\mathcal{R}^3)+{\rm div},
$$
where {\rm div} takes into account total derivatives and $\mathcal{O}(\mathcal{R}^3)$ only contributes at order $\mathcal{O}(h^3).$ Hence, in the linearized regime we can set $\mathcal{F}_3(\Box)=0$ without loss of generality.} $\mathcal{O}(h^2)$ \cite{Biswas:2011ar}
\begin{eqnarray}
\non
S&=&\frac{1}{4}\int \left\lbrace \frac{1}{2}h_{\mu\nu}f(\Box)\Box h^{\mu\nu}-h_{\mu}^{\sigma}f(\Box)\partial_{\sigma}\partial_{\nu}h^{\mu\nu}\right.\\[2mm]
\non
&&+\,h\,g(\Box)\partial_{\mu}\partial_{\nu}h^{\mu\nu} -\frac{1}{2}h\,g(\Box)\Box h\\[2mm]
&&\left.+\,\frac{1}{2}h^{\lambda\sigma}\frac{f(\Box)-g(\Box)}{\Box}\partial_{\lambda}\partial_{\sigma}\partial_{\mu}\partial_{\nu}h^{\mu\nu}\right\rbrace d^4x ,
\label{lin-quad-action}
\end{eqnarray}
where $h\equiv\eta_{\mu\nu}h^{\mu\nu}$ is the trace of the metric perturbation and we have defined
\begin{eqnarray}
f(\Box)&=&1+\frac{1}{2}\mathcal{F}_2(\Box)\Box\,,\\[2mm]
g(\Box)&=&1-2\mathcal{F}_1(\Box)\Box-\frac{1}{2}\mathcal{F}_2(\Box)\Box\,.
\end{eqnarray}
The corresponding linearized field equations are given by
\begin{eqnarray}
\non
2\kappa^2 T_{\mu\nu}&=&f(\Box)\left(\Box h_{\mu\nu}-\partial_{\sigma}\partial_{\nu}h_{\mu}^{\sigma}-\partial_{\sigma}\partial_{\mu}h_{\nu}^{\sigma}\right)\\[2mm]
\non
&&+\,g(\Box)\left(\eta_{\mu\nu}\partial_{\rho}\partial_{\sigma}h^{\rho\sigma}+\partial_{\mu}\partial_{\nu}h-\eta_{\mu\nu}\Box h\right)\\[2mm]
&&+\, \frac{f(\Box)-g(\Box)}{\Box}\partial_{\mu}\partial_{\nu}\partial_{\rho}\partial_{\sigma}h^{\rho\sigma},
\label{lin-field-eq}
\end{eqnarray}
where the stress-energy tensor sourcing the gravitational field is defined by
\begin{equation}
T_{\mu\nu}\simeq -2\frac{\delta S_m}{\delta h^{\mu\nu}}\,,
\end{equation}
with $S_m$ being the matter action.

We are interested in finding the expression for the linearized spacetime metric in presence of a static point-like source:
\begin{equation}
ds^2=(1+2\phi)dt^2-(1-2\psi)(dr^2+r^2d\Omega^2)\,,\label{isotr-metric}
\end{equation}
where $d\Omega=d\theta^2+{\rm sin}^2 \theta d\varphi^2,$ $\phi$ and $\psi$ are the two metric potentials, while the matter sector is described by
\begin{equation}
T_{\mu\nu}=m\delta_{\mu}^0\delta_{\nu}^0\delta^{(3)}(\vec{r})\,.
\end{equation}
By setting $\kappa h_{00}=2\phi$, $\kappa h_{ij}=2\psi\delta_{ij}$, $\kappa h=2(\phi-3\psi)$ and using the assumption of static pressureless source, i.e. $\Box\simeq-\nabla^2$ and $T=\eta_{\rho\sigma}T^{\rho\sigma}\simeq T_{00}\label{key}$, the modified Poisson equations for the two metric potentials read
\begin{eqnarray}
\label{potpot}
\frac{f(f-3g)}{f-2g}\nabla^2\phi(r)&=&8\pi Gm\delta^{(3)}(\vec{r})\,,\\[2mm]
\frac{f(f-3g)}{g}\nabla^2\psi(r)&=&- 8\pi Gm\delta^{(3)}(\vec{r})\,,
\label{field-eq-pot}
\end{eqnarray}
where $f\equiv f(\nabla^2),$ $g\equiv g(\nabla^2)$ are now functions of the Laplace operator.

The two modified Poisson equations~\eqref{potpot} and~\eqref{field-eq-pot} can be solved with the use of the Fourier transform method, by going to momentum space and then anti-transforming back to coordinate space. Thus, we obtain
\begin{eqnarray}
\non
\phi(r)&=& \displaystyle -8\pi Gm\int \frac{1}{k^2}\frac{f-2g}{f(f-3g)}\,e^{i\vec{k}\cdot \vec{r}}\frac{d^3k}{(2\pi)^3}\\[2mm]
\label{pot1}
&=& \displaystyle-\frac{4Gm}{\pi r}\int_0^{\infty}\frac{f-2g}{f(f-3g)}\frac{{\rm sin}(kr)}{k}\hspace{0.4mm}dk\,,\\[2mm]
\non
\psi(r)&=& \displaystyle 8\pi Gm\int \frac{1}{k^2}\frac{g}{f(f-3g)}\,e^{i\vec{k}\cdot \vec{r}}\frac{d^3k}{(2\pi)^3}\\[2mm]
&=& \displaystyle\frac{4Gm}{\pi r}\int_0^{\infty}\frac{g}{f(f-3g)}\frac{{\rm sin}(kr)}{k}\,dk\,,
\label{fourier-pot}
\end{eqnarray}
where $f\equiv f(-k^2)$ and $g\equiv g(-k^2)$ are now functions of the Fourier momentum squared.

As a first check, we can notice that in the case $f=g$ we recover the weak-field limit of Einstein's GR, 
\begin{equation}\label{grl}
f=g=1\,\,\Longrightarrow\,\,\phi(r)=\psi(r)=-\frac{Gm}{r}\,,
\end{equation}
as expected.

%%%%%%%%%%%%%%%%%%%%%%%%%%%%%%%%%%%%%%%%%%%%%%%%%%%%%%
\section{Oscillation phase expression}
\label{ope}
%%%%%%%%%%%%%%%%%%%%%%%%%%%%%%%%%%%%%%%%%%%%%%%%%%%%%%
In this Section, we want to study the form that the covariant oscillation phase acquires when the spacetime is described by several quadratic models of gravity whose action is given by Eq.~(\ref{quad-action}). Specifically, we refer to the phase that appears in the expression of the flavor transition probability~\eqref{probo}
\be\label{sp}
\mathcal{P}_{\al\to\beta}=\mathrm{sin}^2(2\theta)\,\mathrm{sin}^2\lf(\frac{\varphi_{12}}{2}\ri),
\ee
where  now $\varphi_{12}\equiv\varphi_1-\varphi_2$ denotes 
the oscillation phase in curved background, i.e.
\be
\label{om12}
\Phi|\nu_k\rangle=\varphi_k|\nu_k\rangle\,,
\ee 
with $\Phi$ given by Eq.~\eqref{phasep}. 

\subsection{Vacuum oscillations}
\label{vacoscillations}
Let us start by considering only geometric effects (matter effects
will be accommodated later). %We remark that
%these are enough to put in evidence the main point of this work, i.e. 
%the appearance of a SEP violating term in the neutrino oscillation phase.
Since our attention is focused on the analysis of a radial propagation, it is possible to prove that, in all the upcoming discussions, we have $A_{G\mu}=0$.
%\be\label{zer}
%\frac{d x^\mu_{null}}{d\lambda}A_{G\mu}=0\,,
%\ee
%which allows us to treat Eq.~(\ref{eqn}) more easily.
Actually, this is always true for such diagonal metrics as the one in Eq.~\eqref{isotr-metric} (see e.g. Refs. \cite{Cardall,Mukhopadhyay:2005gb,for}). In fact, a brief analysis of Eq. \eqref{eqn:A} shows that a non-vanishing $A_{G\mu}$ requires non-zero off-diagonal components of the tetrads. It is immediate to verify that, in our case,
\be\label{tetr}
e_{\hat{0}}^0=1-\phi\,, \qquad e_{\hat{j}}^i=\lf(1+\psi\ri)\delta^{i}_{j}\,.
\ee
At this point, the phase $\varphi_{12}$ takes the form \cite{Cardall,for}
\be\label{ph}
\varphi_{12}=\frac{\Delta m^2}{2}\int_{\lambda_{A}}^{\lambda_B}d\lambda=\frac{\Delta m^2}{2}\int_{\ell_A}^{\ell_B}\frac{d\ell}{E_\ell}\,,
\ee  
where we have made use of Eq.~(\ref{line}) in the second step and $E_\ell=\frac{}{}e_{\hat{0}}^0 E$ is the energy measured by a {\it locally} inertial observer momentarily at rest in the curved spacetime and $E$ represents the energy measured by an inertial observer at rest at infinity. Since we are assuming to work with a stationary metric, it is worth emphasizing that $E\equiv P_0$ is a conserved quantity.

By use of Eq.~(\ref{tetr}), Eq.~(\ref{ph}) can be rephrased as
\be\label{ph2}
\varphi_{12}\,=\,\frac{\Delta m^2}{2E}\int_{r_A}^{r_B}\lf[1+\phi(r)-\psi(r)\ri]d r\,,
\ee
given that $d\ell^2=\lf(1-2\psi\ri)d r^2$ for radial motion.

In accordance with the reasoning exhibited so far, the flavor oscillation probability can be rewritten as
\begin{eqnarray}
\label{sp2}
\non
\mathcal{P}_{\al\to\beta}&\hspace{-1mm}=\hspace{-1mm}&\displaystyle \mathrm{sin}^2(2\theta)\,\mathrm{sin}^2\left\lbrace \frac{\Delta m^2}{4 E}\int_{r_A}^{r_B}\lf[1+\phi(r)-\psi(r)\ri]d r\right\rbrace \\[2mm]
&\hspace{-1mm}=\hspace{-1mm}&\mathrm{sin}^2(2\theta)\,\mathrm{sin}^2\lf[\frac{\Delta m^2}{4 E}(r_B-r_A)\,+\, \frac{\ph_{_{SEP}}}{2}\ri], 
\end{eqnarray}
where we have introduced the shorthand notation
\be \label{phsep}
\ph_{_{SEP}} \, = \, \frac{\Delta m^2}{2 E}\int_{r_A}^{r_B}\lf[\phi(r)-\psi(r)\ri]\,dr \, , 
\ee
whose meaning will be clarified in the next subsection.

Depending on the choice of the form factors in Eq.~(\ref{quad-action}), we expect $\mathcal{P}_{\al\to\beta}$ to be a function of the free parameters of the selected quadratic theory of gravity. In turn, this implies that the neutrino oscillation probability is strictly related to the model used to investigate the geometric features of the curved background. 

In addition, it is possible to show that the covariant oscillation phase can always be split in three different contributions. Guided by this idea, one can check that Eq.~(\ref{ph2}) always includes the following terms:
\be\label{phs}
\varphi_{12}=\varphi_{0}+\varphi_{_{GR}}+\varphi_{_{Q}}\,,
\ee
where $\varphi_{_0}$ is formally the same as the usual ``flat'' phase when $m=0$~(\ref{minkp}), $\varphi_{_{GR}}$ is the quantity associated to GR, whereas $\varphi_{_{Q}}$ includes all the corrections  due to the quadratic models of gravity. The feasibility of such a procedure is guaranteed by the fact that the two metric potentials $\phi$ and $\psi$ can be always recast as $\phi=\phi_{_{GR}}+\phi_{_{Q}}$ and $\psi=\psi_{_{GR}}+\psi_{_{Q}}=\phi_{_{GR}}+\psi_{_{Q}}$, respectively (since $\psi_{_{GR}}=\phi_{_{GR}}$, as seen in the previous Section). At this point, the appearance of $\varphi_{_0}$ ensues from a simple consideration: starting from Eq.~(\ref{ph2}), indeed, we can cast $E$ in terms of the local energy by using $E_\ell=\lf(1-\phi\ri)E$ and then introduce the proper distance covered by the neutrino propagating on a curved background:
\be\label{lp}
L_p=\int_{r_A}^{r_B}\sqrt{-g_{rr}}\,dr=\int_{r_A}^{r_B}\lf[1-\psi(r)\ri]\,dr\,.
\ee
In view of these notions, the covariant phase~(\ref{ph2}) can be expressed as
\be\label{ph3}
\varphi_{12}=\frac{\Delta m^2L_p}{2E_\ell}\lf[1-\phi(r_B)+\frac{1}{L_p}\int_{r_A}^{r_B}\hspace{-2mm}\phi(r)\,dr\ri].
\ee
Hence, the first term on the r.h.s. precisely returns the phase in Eq.~(\ref{minkp}), with the difference that here it is written as a function of the local energy and the proper propagation distance. Since we are interested in a slightly curved background (i.e. in the weak-field regime), 
we will now report the explicit linearized expressions for 
$\varphi_{_{GR}}$~\cite{for}
\be\label{grp}
\varphi_{_{GR}}=\frac{\Delta m^2L_p}{2E_\ell}\lf[\frac{Gm}{r_B}-\frac{Gm}{L_p}\ln\lf(\frac{r_B}{r_A}\ri)\ri],
\ee
while the contribution to the phase only due to the quadratic theories correction is:
\bea\label{qphase}
\varphi_{_{Q}} \, = \, \frac{\Delta m^2 L_p}{2E_\ell}\lf[\frac{1}{L_p}\int_{r_A}^{r_B}\hspace{-2mm}\phi_{_{Q}}(r)\,dr-\phi_{_{Q}}(r_B) \ri] .
\eea

Let us remark that we are working in the linear regime, where we can always perform analytical computations. Interesting physical scenarios in which our analysis and outcomes may be tested are the ones of solar and supernova neutrinos, the latter being relevant due to the extremely large fluxes of particles produced in a wide range of energies. 
A detailed study of these aspects, however, goes beyond the scope of the present manuscript and will be treated elsewhere. We remand to the existing literature for more specific discussions on this (see, for instance, Refs.~\cite{Capozziello:1999qm,relevance, Grossman} and therein). 
Furthermore, we emphasize that, consistently with our work assumption, the term $|\varphi_0|$ will be always larger than the gravitational corrections, indeed we can have the following two cases: $|\varphi_{_0}|>|\varphi_{_{GR}}|\gtrsim|\varphi_{_{Q}}|$ and $|\varphi_{_0}|>|\varphi_{_{Q}}|\gtrsim|\varphi_{_{GR}}|,$ which are both compatible with the linearized approximation. Given such inequalities and by making a comparison with experiments, one can eventually put constraints on the free parameters of the given gravitational theory.
%%%%%%%%%%%%%%%%%%%%%%%%%%%%%%%%%%%%%%%%%%%%%%%%%%%%%%%%%%%%%%%%%%%%%%%%%%%%%%%%%%%%%%%%%%%%%%%%%%%%%%%%%
\subsection{A link with the equivalence principle violation}
\label{Newsec}
Before applying the aforementioned considerations to several quadratic theories of gravity, it is worth focusing the attention on a possible connection between the covariant phase~(\ref{sp2}) and the violation of the strong equivalence principle~\cite{will}. In particular, by looking at Eq.~(\ref{sp2}), we refer to the term proportional to $\phi-\psi$, which in the case of pure GR would be identically zero. However, this difference can be recognized as a clear signal for SEP violation, since the two metric potentials are not equal~\cite{will,nord}. 

In view of the last consideration, one can indeed evaluate the Eddington-Robertson-Schiff parameter $\ga$ that arises from a post-Newtonian limit and which is related to how much space-curvature is produced by unit rest mass of the gravitational source (for a detailed review of this topic see Refs.~\cite{will,sep}). If we adopt the metric~(\ref{isotr-metric}), one can show that
\be\label{ppn}
\ga=\frac{\psi}{\phi}\,,
\ee
but since we have already pointed out that both metric potentials can be decomposed in a term related to GR and a correction due to the presence of quadratic contributions in the gravitational action, the previous equation can be also cast into the (more convenient) form
\be\label{ppn2}
\ga-1=\frac{\psi_{_{Q}}-\phi_{_{Q}}}{\phi}\,.
\ee
As expected, if the gravitational action is the Einstein-Hilbert one, we have $\phi_{_{Q}}=\psi_{_{Q}}=0$, which means $\ga=1$, that is the known value of such a parameter in the case of GR.

At this point, in order to properly quantify the violation of SEP, it is customary to analyze the so-called Nordtvedt parameter $\eta$~\cite{par,par2}, defined as
\be\label{nopa}
\eta=4\lf(\beta-1\ri)-\lf(\ga-1\ri),
\ee
where the post-Newtonian parameter $\beta$ quantifies nonlinear gravitational effects. Strong equivalence principle is violated as long as $\eta\neq 0$~\cite{par2}. 

In reporting the expression~(\ref{nopa}), we have tacitly required the absence of anisotropies and preferred-frame effects~\cite{will,sep}, which should have been described by other post-Newtonian parameters that have been set to zero in the current analysis (see Ref.~\cite{par2} for more details). If we perform the further assumption that nonlinear effects are essentially described by the contributions coming from GR, then $\beta=1$~\cite{will}, which in turn entails
\be\label{nopa2}
\eta=1-\ga\,.
\ee
In principle, for a given metric as in Eq.~(\ref{isotr-metric}), the quantity $\ga$ depends on the position, namely $\ga\equiv\ga\lf(r\ri)$. However, we can assume to investigate the scenario in which $\ga$ is slowly varying with respect to the spatial coordinates\footnote{In other words, we can restrict the attention to the spatial region in which variations of $\ga$ are negligible.}. Therefore, we may treat it as a constant, in such a way to render all the considerations centered around $\eta$ enforceable. Indeed, the analysis performed on the SEP violation with the aid of the Nordtvedt parameter has been developed by taking the post-Newtonian expansion coefficients to be constant.

Now, we can observe from Eq.~(\ref{ppn2}) that the deviation from the GR prediction is strictly related to the difference of the metric potentials associated to the quadratic part of the gravitational action. By means of Eq.~(\ref{nopa2}), such a discrepancy is an evident indication of the SEP violation. Hence, from Eq.~(\ref{sp2}), it follows that the neutrino oscillation phase does {\it discriminate} whether the particle propagates in the conditions in which SEP is satisfied or not. This explains the meaning
of $\varphi_{_{SEP}}$ in Eq.~\eqref{phsep}.

The reason for the occurrence of SEP violation could be readily attributed to the emergence of a 
non-standard term in the Dirac Hamiltonian that explicitly depends upon the difference $\phi-\psi;$ this aspect will be investigated in future works.

\subsection{Matter effects}
Let us now discuss how to generalize our previous considerations when effects of background matter are taking into account as shown in Section \ref{matef}.

Neutrino flavor states evolution can be described as~\cite{Cardall}
\begin{equation}
\label{evmat}  
|\nu_f (\la_B)\ran =  \exp\left[{\int^{\la_{B}}_{\la_A} \left(\frac{M^2_f}{2}\, - \,\frac{d x^\mu_{null}}{d \lambda}A_{f\mu} P_L\right) d \la}\right] |\nu_f \ran \, , 
\end{equation}
where
\begin{equation}
|\nu_f(\la)\ran \,=\,  \begin{pmatrix} |\nu_e(\la)\ran \\[1mm] |\nu_\mu(\la)\ran \end{pmatrix},
\end{equation}
is the doublet of spinors in the flavor basis and $|\nu_\al\ran = |\nu_\al(\la_A)\ran$ ($\alpha=e,\mu$) is
defined as in Eq. \eqref{flavneutr}. Here we have taken into account that 
the gravitational corrections
in the present case are vanishing, i,e $A_{G\mu}=0$. 

By assuming that the experimental setup is at rest with respect to the electron background, we get 
\begin{equation}
\label{evmat1} \, 
|\nu_f (\la)\ran \, = \, \exp\left[{\frac{1}{2E} \, \int^{r_B}_{r_A}\frac{\widetilde{M}^2_f}{2}\, d r}\right]|\nu_f \ran \, , 
\end{equation}
with 
\bea
\label{MTILDE}
\widetilde{M}^2_{f} \, \equiv \, M_f^2 (1+\phi-\psi) \, + \, V_f \, (1-\psi) \, ,  
\eea
where $V_f$ is defined by
\bea \label{vf}
V_{f} \, \equiv \, \begin{pmatrix} v(r) & \,\,0 \\ 0 & \,\,0  \end{pmatrix},
\eea
and $v(r)=2 \sqrt{2}\hspace{0.2mm}E\hspace{0.2mm}G_F\hspace{0.2mm}n_e(r)\hspace{0.2mm}P_L$.

We now diagonalize the matrix~\eqref{MTILDE} via the transformation
\be
\widetilde{M}^2\,=\,\begin{bmatrix} \widetilde{m}_1^2 & 0 \\ 0 & \widetilde{m}_2^2 \end{bmatrix} \, \equiv \, U^\dag(\widetilde{\theta}) \, \widetilde{M}^2_f \, U(\widetilde{\theta}) \, ,
\ee
where 
\bea
\tan 2 \widetilde{\theta} \,=\,\frac{\Delta m^2\left[\Delta m^2\cos2\theta-v(1+\phi)\right]\sin2\theta}{{\left(\Delta m^2\cos2\theta-v\right)}^2}\,.
\eea
By following Ref.~\cite{Cardall}, we 
assume the resonance condition $\widetilde{\theta}=\pi/4$, which leads to
\be
\label{rc}
v(r)\,\equiv\,v\,=\,\Delta m^2\cos2\theta\,.
\ee
This means that the interaction between the electron background
and gravity is negligible in the weak field approximation, thus yielding
a constant value of $v$.

Equation~\eqref{evmat1} can be rewritten as:
\begin{equation}
\label{evmat2} 
|\nu_f (\la)\ran \, = \, U(\widetilde{\theta}) \, \exp\left[{\frac{1}{2E} \, \int^{r_B}_{r_A} \, \frac{\widetilde{M}^2}{2}\, d r}\right] |\widetilde{\nu}_m \ran \, , 
\end{equation}
where
\be
|\widetilde{\nu}_m\ran\,=\, \begin{pmatrix} |\widetilde{\nu}_1\ran \\[1mm] |\widetilde{\nu}_2\ran \end{pmatrix}  \,\equiv\, U^\dag(\widetilde{\theta}) \, |\nu_f\ran\,.
\ee
Therefore, the flavor oscillation probability takes the form
\begin{eqnarray}
\mathcal{P}_{\al\to\beta}&\hspace{-1mm}=\hspace{-1mm}&\displaystyle\mathrm{sin}^2 \, \lf(\frac{\widetilde{\varphi}_{12}}{2}\ri), 
\end{eqnarray}
where we have exploited the condition $\sin^2(2\widetilde\theta)=1$ and
\be\label{phti}
\widetilde{\varphi}_{12}\, = \, \frac{\Delta \mu^2}{2E}\int_{r_A}^{r_B}\lf[1+\phi(r)-\psi(r)+\chi \,\phi(r)\ri]d r\,.
\ee
Here we have used the following notation
\bea
\chi & \equiv & \frac{\Delta m^2 \, (\Delta m^2+v \cos 2 \theta)}{\lf(\Delta \mu^2  \ri)^2}-1 \, , \\[2mm]
\Delta \mu^2 & \equiv & \lf(\Delta \widetilde{m}^2 \ri)_{\phi=\psi=0} \non \\[2mm] 
& = & \sqrt{\lf(\Delta m^2\ri)^2 \, + \, v^2 - 2 \, \Delta m^2 \,  v \, \cos 2 \theta} \, .
\eea

Now, as in the case of vacuum oscillations, $\widetilde{\varphi}_{12}$  can be expanded as
\be\label{phti1}
\widetilde{\varphi}_{12}\, = \, \frac{\Delta \mu^2}{2E}\lf(r_B-r_A \ri)+\frac{\chi\Delta \mu^2}{2E}\int_{r_A}^{r_B} \phi(r) \, d r\,+\widetilde{\varphi}_{_{SEP}}\,,
\ee
where we have defined
\be
\widetilde{\varphi}_{_{SEP}} \, \equiv \, \frac{\Delta \mu^2}{2E} \int^{r_B}_{r_A} \, \lf[\phi(r) -\psi(r)\ri] d r \, .
\ee
In terms of local quantities, Eq.~\eqref{phti1} becomes 
\be\label{ph3mat}
\widetilde\varphi_{12}\,=\,\frac{\Delta \mu^2L_p}{2E_\ell}\lf[1-\phi(r_B)+\frac{(1+\chi)}{L_p}\int_{r_A}^{r_B}\hspace{-2mm}\phi(r)\,dr\ri].
\ee
Note that, when matter effects are negligible or absent ($v \rightarrow 0$), then $\De \mu^2 \to \De m^2$ and $\chi \rightarrow 0$, thus recovering Eq.~\eqref{ph3}. Let us also remark that considerations on the possibility of constraining the free parameters of extended theories can be easily repeated as in the case of vacuum oscillations.

%%%%%%%%%%%%%%%%%%%%%%%%%%%%%%%%%%%%%%%%%%%%%%%%%%%%%%%%%%%%%%%%%%%%%%%%%%%%%%%%%%%%%%%

\section{Applications}\label{example}

To better explore the above scenario, in the following we determine $\varphi_{_{Q}}$ and $\varphi_{_{SEP}}$ appearing in the oscillation formula for several quadratic theories whose relevance has been pointed out in the recent literature. For simplicity, we only compute the oscillation phases in vacuum as their generalization to the presence of matter is straightforward. Indeed, for each gravitational theory we consider below, once the vacuum formula \eqref{ph3} for the phase is known, one can obtain the corresponding expression including matter effects by making the following substitutions\footnote{We recall that the following analysis is carried out
by assuming the resonance condition~\eqref{rc}.}:
\begin{eqnarray}
\label{mat}
\Delta m^2 &\rightarrow&\Delta\mu^2,\\[3mm] 
\frac{1}{L_p}\int^{r_B}_{r_A}\phi(r)\,dr & \rightarrow& \frac{1+\chi}{L_p}\int^{r_B}_{r_A}\phi(r)\,dr\,.
\label{mat2}
\end{eqnarray}
Before going any further, it is worthwhile mentioning that all the theories we will study are characterized by new physical scales which are described by free parameters. The best constraints on such parameters come from torsion balance experiments \cite{Kapner:2006si} with which Newton's law has been tested up to roughly $10$ micrometers.

\subsection{$f(\mathcal{R})$-gravity}

We first address the easiest extension of the Einstein-Hilbert action by including a Ricci squared contribution with a constant form-factor $\alpha$
\begin{equation}
\mathcal{F}_1=\alpha\,,\,\,\,\mathcal{F}_2=0\,\,\Longrightarrow\,\,f=1\,,\,\,\,g=1-2\,\alpha\,\Box\,.
\end{equation}
This choice belongs to the class of $f(\mathcal{R})$ theories, where the Lagrangian is truncated up to the order $\mathcal{O}(\mathcal{R}^2)$
\begin{equation}
f(\mathcal{R})\simeq \mathcal{R}+\frac{\alpha}{2} \mathcal{R}^2,
\end{equation}
and the cosmological constant is set to zero.

For the above selection of the form-factors, the two metric potentials in Eqs.~\eqref{pot1} and~\eqref{fourier-pot} become
\begin{eqnarray}\label{fr}
\phi(r)&=&\displaystyle -\frac{Gm}{r}\left(1+\frac{1}{3}e^{-m_0r}\right),\\[2mm]
\psi(r)&=&\displaystyle -\frac{Gm}{r}\left(1-\frac{1}{3}e^{-m_0r}\right),
\end{eqnarray}
where $m_0=1/\sqrt{3\,\alpha}$ is the mass of the spin-$0$ massive degree of freedom coming from the Ricci scalar squared contribution.

The Eddington-Robertson-Schiff parameter $\ga$ for this model turns out to be
\be\label{g1}
\ga=\frac{1-\frac{1}{3}e^{-m_0r}}{1+\frac{1}{3}e^{-m_0r}}\simeq 1-\frac{2}{3}e^{-m_0r}\,,
\ee
where after the second equality we have performed an expansion for small values of the exponential function correction. Such a limit is feasible because we expect $m_0$ to be large. Note that the GR limit (and therefore $\ga=1$) is restored for $m_0\to\infty$.

By using $\phi=\phi_{_{GR}}+\phi_{_{Q}}$, with
\be\label{q1}
\phi_{_Q}(r)=-\frac{1}{3}\frac{Gm}{r}e^{-m_0r}\,,
\ee
and relying on Eq.~(\ref{qphase}), we obtain
\be\label{q11}
\varphi_{_{Q}}=\frac{\Delta m^2L_p}{2E_\ell}\left\{\frac{Gm\,e^{-m_0 r_B}}{3 r_B}-\frac{Gm}{3L_p}\Bigl[\mathrm{Ei}\lf(-m_0 r\ri)\Bigr]^{r_B}_{r_A}\right\} \, ,
\ee
where the special function
\be
\mathrm{Ei}(x)=-\int_{-x}^{\infty}\frac{e^{-\xi}}{\xi}\,d\xi\,,
\ee
is known as the exponential integral function~\cite{grad} and we have introduced the shorthand notation
\be\label{sh}
\Bigl[f(x)\Bigr]^{x_B}_{x_A}\equiv f(x_B)-f(x_A)\,.
\ee
%Consequently, Eq.~(\ref{cons}) applied to this model leads to the following bound:
%\be\label{b1}
%\lf|\frac{e^{-m_0\lf(L_p+r_A\ri)}}{L_p+r_A}-\frac{1}{L_p}\Bigl[\mathrm{Ei}\lf(-m_0r\ri)\Bigr]^{L_p+r_A}_{r_A}\ri|<\frac{3}{Gm}\,.
%\ee
Moreover, from Eq.~\eqref{phsep}, one can evaluate the SEP violating phase as follows:
\bea\label{sep1}
\ph_{_{SEP}}\, = \, \frac{\De m^2 \, G m}{3 E_\ell} \Bigl[\mathrm{Ei}\lf(-m_0 r\ri)\Bigr]^{r_B}_{r_A}\,.
\eea
This term can be identified with the second contribution in the r.h.s. of Eq.~\eqref{q11}.

\subsection{Stelle's fourth-order gravity}

Let us now consider Stelle's fourth-order gravity~\cite{Stelle}, which is achieved with the following form-factors:
\begin{equation}
\mathcal{F}_1=\alpha\,,\,\mathcal{F}_2=\beta\,\Longrightarrow\,f=1+\frac{1}{2}\beta\,\Box,\,\,g=1-2\alpha\,\Box-\frac{1}{2}\beta\,\Box\,.
\end{equation}
Unlike the $f(\mathcal{R})$ case, the Ricci tensor squared contribution in the action is clearly recognizable through a constant, non-vanishing form-factor. It is possible to check that the gravitational action related to this model turns out to be renormalizable~\cite{Stelle}. 

For the above choice of the form-factors, the two metric potentials in Eqs.~\eqref{pot1} and~\eqref{fourier-pot} now read
\begin{eqnarray}
\phi(r)&=& -\frac{Gm}{r}\left(1+\frac{1}{3}e^{-m_0r}-\frac{4}{3}e^{-m_2r}\right),\\[2mm]
\psi(r)&=& -\frac{Gm}{r}\left(1-\frac{1}{3}e^{-m_0r}-\frac{2}{3}e^{-m_2r}\right),\label{stelle-pot}
\end{eqnarray}
where $m_0=2/\sqrt{12\,\alpha+\beta}$ and $m_2=\sqrt{2/(-\beta)}$ correspond to the masses of the spin-$0$ and of the spin-$2$ massive mode, respectively. 
In order to avoid tachyonic solutions, we need to require $\beta<0$. Additionally, the spin-$2$ mode is a ghost-like degree of freedom. Such an outcome is not surprising, since it is known that, for any local higher derivative theory of gravity, ghost-like degrees of freedom always appear\footnote{See Refs.~\cite{Anselmi,Anselmi:2018ibi} for recent works in which the authors have shown that both renormalizability and unitarity can be made to coexist by implementing a new quantization prescription.}.

The factor $\ga$ appearing in Eq.~(\ref{ppn}) for Stelle's fourth-order gravity is given by
\be\label{g2}
\ga=\frac{1-\frac{1}{3}e^{-m_0r}-\frac{2}{3}e^{-m_2r}}{1+\frac{1}{3}e^{-m_0r}-\frac{4}{3}e^{-m_2r}}\simeq 1-\frac{2}{3}e^{-m_0r}+\frac{2}{3} e^{-m_2r}\,.
\ee
As for the previous case, the limit of large masses $m_0,m_2\to\infty$ returns GR.

If we single out the contribution of this quadratic model to the potential $\phi$, we note that
\be\label{q2}
\phi_{_{Q}}(r)=-\frac{1}{3}\frac{Gm}{r}e^{-m_0r}+\frac{4}{3}\frac{Gm}{r}e^{-m_2r}\,.
\ee
Hence, the phase $\varphi_{_{Q}}$ turns out to be
\bea\label{q22}\non
\varphi_{_{Q}}&=&\frac{\Delta m^2L_p}{2E_\ell}\Bigl\{\frac{Gm\,e^{-m_0 r_B}}{3 r_B}-\frac{4Gm\,e^{-m_2 r_B}}{3 r_B}\\[2mm]
&&\hspace{-12mm}-\frac{Gm}{3L_p}\Bigl[\mathrm{Ei}\lf(-m_0r\ri)\Bigr]^{r_B}_{r_A}\hspace{-1.5mm}+\frac{4Gm}{3L_p}\Bigl[\mathrm{Ei}\lf(-m_2r\ri)\Bigr]^{r_B}_{r_A}\Bigr\} \, .
\eea
The SEP violating phase \eqref{phsep} is now
\bea\label{sep2}
\ph_{_{SEP}}\, = \, \frac{\De m^2 \, G m}{3 E_\ell} \Bigl[\mathrm{Ei}\lf(-m_0 r\ri)-\mathrm{Ei}\lf(-m_2 r\ri)\Bigr]^{r_B}_{r_A}\,.
\eea
%
%Eq.\eqref{q2} can be thus rewritten as
%\bea \non
%\varphi_{_{Q}}&=&\frac{\Delta m^2L_p}{2E_\ell}\Bigl\{\frac{Gm\,e^{-m_0 r_B}}{3 r_B}-\frac{4Gm\,e^{-m_2 r_B}}{3 r_B}\\[2mm]
%&+&\frac{Gm}{3L_p}\Bigl[\mathrm{Ei}\lf(-m_2r\ri)\Bigr]^{r_B}_{r_A}\Bigr\}-\ph_{_{SEP}} \, .
%\eea
%

%which in turn implies
%\bea\label{b2}
%&&\hspace{-5mm}\Bigl|\frac{e^{-m_0\lf(L_p+r_A\ri)}-4e^{-m_2\lf(L_p+r_A\ri)}}{L_p+r_A}\\[2mm]\non
%&&\hspace{-5mm}-\frac{1}{L_p}\Bigl\{\Bigl[\mathrm{Ei}\lf(-m_0r\ri)\Bigr]^{L_p+r_A}_{r_A}\hspace{-1.5mm}-4\Bigl[\mathrm{Ei}\lf(-m_2r\ri)\Bigr]^{L_p+r_A}_{r_A}\Bigr\}\Bigr|<\frac{3}{Gm}.
%\eea

\subsection{Sixth-order gravity}
Let us now deal with a sixth-order gravity model, which is an example of super-renormalizable theory~\cite{Giacchini:2018gxp,Accioly:2016qeb}
\begin{eqnarray}
\non
&&\hspace{-5mm}\mathcal{F}_1=\alpha\,\Box\,,\,\,\mathcal{F}_2=\beta\,\Box\\[2mm]
&&\hspace{-5mm}\Longrightarrow\,f=1+\frac{1}{2}\beta\,\Box^2,\,\,\,g=1-2\alpha\Box^2-\frac{1}{2}\beta \Box^2\,.
\end{eqnarray}
It is possible to show that the two metric potentials in Eqs.~\eqref{pot1} and~\eqref{fourier-pot} assume the following expressions:
\begin{eqnarray}
\non
\phi&\hspace{-1mm}=\hspace{-1mm}& -\frac{Gm}{r}\!\left(1+\frac{1}{3}e^{-m_0r}\,{\rm cos}(m_0r)-\frac{4}{3}e^{-m_2r}\,{\rm cos}(m_2r)\right),\\[2mm]\\\non
\psi&\hspace{-1mm}=\hspace{-1mm}&-\frac{Gm}{r}\!\left(1-\frac{1}{3}e^{-m_0r}\,{\rm cos}(m_0r)-\frac{2}{3}e^{-m_2r}\,{\rm cos}(m_2r)\right),\\\label{sixth-pot}
\end{eqnarray}
where the masses of the spin-$0$ and spin-$2$ degrees of freedom are now given by $m_0=2^{-1/2}(-3\,\alpha-\beta)^{-1/4}$ and $m_2=(2\,\beta)^{-1/4}$, respectively.
Note that, in this case, tachyonic solutions are avoided for $-3\,\alpha-\,\beta>0$, which can be satisfied by the requirement $\alpha<0$ and $-3\,\alpha>\,\beta$, with $\beta>0$. 
The current higher derivative theory of gravity has no real ghost-modes around the Minkowski background, but a pair of complex conjugate poles with equal real and imaginary parts~\cite{Accioly:2016qeb}, and corresponds to the so called Lee-Wick gravity~\cite{Modesto:2015ozb}. It is worthwhile noting that in this model the unitarity condition is not violated, indeed the optical theorem still holds~\cite{Anselmi,Anselmi:2017yux,Anselmi:2017lia}.

The parameter $\ga$ related to SEP violation now reads
\bea\label{g3}
\non
\hspace{-4mm}\ga&=&\frac{1-\frac{1}{3}e^{-m_0r}\cos\lf(m_0r\ri)-\frac{2}{3}e^{-m_2r}\cos\lf(m_2r\ri)}{1+\frac{1}{3}e^{-m_0r}\cos\lf(m_0r\ri)-\frac{4}{3}e^{-m_2r}\cos\lf(m_2r\ri)}\\[2mm]
\hspace{-8mm}&\simeq& 1-\frac{2}{3}e^{-m_0r}\cos^2\lf(m_0r\ri)+\frac{2}{3} e^{-m_2r}\cos^2\lf(m_2r\ri).
\eea
%with the exponentials treated as small quantities by virtue of the same considerations contained in the previous subsection.
For this model, we have 
\be\label{q3}
\phi_{_{Q}}(r)=-\frac{1}{3}\frac{Gm}{r}e^{-m_0r}\cos(m_0r)\,+\,\frac{4}{3}\frac{Gm}{r}e^{-m_2r}\cos(m_2r)\,.
\ee
Accordingly, the gravitational phase due to the quadratic part of the action reads
\bea\label{q33}\non
\varphi_{_{Q}}&=&\frac{\Delta m^2L_p}{2E_\ell}\Bigl\{\frac{Gm\,e^{-m_0\, r_A}}{3 r_B}\cos\lf[m_0 r_B\ri]\\[2mm]\non
&&-\frac{4Gm\,e^{-m_2 r_B}}{3 r_B}\cos\lf[m_2 r_B\ri]\\[2mm]\non
&&-\frac{Gm}{6L_p}\Bigl[\mathrm{Ei}\lf(k_1m_0r\ri)+\mathrm{Ei}\lf(k_2m_0r\ri)\Bigr]^{r_B}_{r_A}\\[2mm]
&&+\frac{2Gm}{3L_p}\Bigl[\mathrm{Ei}\lf(k_1m_2r\ri)+\mathrm{Ei}\lf(k_2m_2r\ri)\Bigr]^{r_B}_{r_A}\Bigr\}\,,
\eea
with
\be
k_1=-1-i, \qquad k_2=-1+i\,.
\ee
The SEP violating phase is now
\bea\label{sep3}
\varphi_{_{SEP}}&=&\frac{\De m^2 \, Gm}{3 E_\ell}\Bigl\{\Bigl[\mathrm{Ei}\lf(k_1m_2r\ri)+\mathrm{Ei}\lf(k_2m_2r\ri)\Bigr]^{r_B}_{r_A} \non \\[2mm]
&-&\Bigl[\mathrm{Ei}\lf(k_1m_0r\ri)+\mathrm{Ei}\lf(k_2m_0r\ri)\Bigr]^{r_B}_{r_A}\Bigr\}\,.
\eea
%The application of Eq.~(\ref{cons}) then returns
%\bea\label{b3}
%&&\Bigl|\frac{e^{-m_0\lf(L_p+r_A\ri)}\cos\lf[m_0\lf(L_p+r_A\ri)\ri]}{L_p+r_A}\\[2mm]\non
%&&\hspace{5mm}-\frac{e^{-m_2\lf(L_p+r_A\ri)}\cos\lf[m_2\lf(L_p+r_A\ri)\ri]}{L_p+r_A}\\[2mm]\non
%&&\hspace{5mm}-\frac{1}{L_p}\Bigl[\mathrm{Ei}\lf(k_1m_0r\ri)+\mathrm{Ei}\lf(k_2m_0r\ri)\Bigr]^{L_p+r_A}_{r_A}\\[2mm]\non
%&&\hspace{5mm}+\frac{4}{L_p}\Bigl[\mathrm{Ei}\lf(k_1m_2r\ri)+\mathrm{Ei}\lf(k_2m_2r\ri)\Bigr]^{L_p+r_A}_{r_A}\Bigr|<\frac{3}{Gm}.
%\eea

\subsection{Ghost-free infinite derivative gravity}
We now consider an example of ghost-free non-local theory of gravity~\cite{Biswas:2005qr,Modesto:2011kw,Biswas:2011ar,Biswas:2013cha,Biswas:2016etb,Edholm:2016hbt,Koshelev:2017bxd,Buoninfante:2018xiw,Buoninfante:2018rlq,Buoninfante:2018stt,Buoninfante:2018mre,Buoninfante:2018xif,Mazumdar:2018xjz,Buoninfante:2018gce,Buoninfante:2019swn}. For the sake of clarity, we  adopt the simplest ghost-free choice for the non-local form-factors~\cite{Biswas:2011ar}
\begin{equation}
\mathcal{F}_1=-\frac{1}{2}\mathcal{F}_2=\,\frac{1-e^{\Box/M_s^2}}{2\,\Box}\,\,\Longrightarrow\,\,f=g=e^{\Box/M_s^2}, \label{ghost-free-choice}
\end{equation}
where $M_s$ is the scale at which the non-locality of the gravitational interaction should become manifest. Note that, for the special ghost-free choice in Eq.~\eqref{ghost-free-choice}, {\it no} extra degrees of freedom other than the massless transverse spin-$2$ graviton propagate around the Minkowski background. 

Since we have chosen $f=g$, the metric potentials of Eqs.~\eqref{pot1} and~\eqref{fourier-pot} coincide
\begin{equation}
\phi(r)=\psi(r)=-\frac{Gm}{r}\,{\rm Erf}\left(\frac{M_sr}{2}\right),\label{ghost-free-pot}
\end{equation}
where
\be\label{erf}
\mathrm{Erf}(x)=\frac{2}{\sqrt{\pi}}\int^x_0e^{-t^2}dt\,,
\ee
is the error function~\cite{grad}.

Note that, since the metric potentials are equal, we automatically obtain $\ga=1$ as in GR, which means that, at least from our study, there is no additional contribution to the neutrino oscillation phase that can be directly related to SEP violation. Indeed, because of our assumptions, for this theory it is straightforward to check that the reduced Nordtvedt parameter in Eq. \eqref{nopa2} identically vanishes. 

By introducing the complementary error function~\cite{grad},
\be\label{erfc}
\mathrm{Erfc}(x)=1-\mathrm{Erf}(x)\,,
\ee
one can prove that
\be\label{q4}
\phi_{_{Q}}(r)=\frac{Gm}{r}\,\mathrm{Erfc}\lf(\frac{M_sr}{2}\ri)\,.
\ee
Therefore, the phase associated to this quadratic model is equal to
\bea\label{q44}
\hspace{-2mm}\varphi_{_{Q}}& = &\frac{\Delta m^2L_p}{2E_\ell}\Bigl\{-\frac{Gm}{r_B}\mathrm{Erfc}\lf[\frac{M_s r_B}{2}\ri]+\frac{Gm}{L_p}\ln\lf(\frac{r_B}{r_A}\ri) \non\\[2mm]
&&\hspace{-2mm}- \frac{Gm}{L_p} \Bigl[\frac{M_sr}{\sqrt{\pi}}{}_2F_2\lf(\frac{1}{2},\frac{1}{2};\frac{3}{2},\frac{3}{2};-\frac{M^2_sr^2}{4}\ri)\Bigr]^{r_B}_{r_A}\Bigr\},
\eea
where we have employed the generalized hypergeometric function~\cite{grad}
\be
{}_pF_q\lf(a_1,\dots,a_p;b_1,\dots,b_q;z\ri)=\sum_{n=0}^{\infty}\frac{(a_1)_n\dots(a_p)_n}{(b_1)_n\dots(b_q)_n}\frac{z^n}{n!}\,,
\ee
with $(x)_n$ being the Pochhammer symbol~\cite{grad}
\be
(x)_0=1, \quad (x)_n=x(x+1)(x+2)\dots(x+n-1)\,.
\ee
%The bound we can derive from Eq.~(\ref{q44}) is thus given by
%\bea\label{b4}
%&&\hspace{-7mm}\Bigl|-\frac{1}{L_p+r_A}\mathrm{Erfc}\lf[\frac{M_s\lf(L_p+r_A\ri)}{2}\ri]+\frac{1}{L_p}\Bigl[\ln\lf(\frac{M_sr}{2}\ri)\\[2mm]\non
%&&\hspace{-7mm}-\frac{M_sr}{\sqrt{\pi}}{}_2F_2\lf(\frac{1}{2},\frac{1}{2};\frac{3}{2},\frac{3}{2};-\frac{M^2_sr^2}{4}\ri)\Bigr]^{L_p+r_A}_{r_A}\Bigr|<\frac{1}{Gm}\,.
%\eea
%Since for this model $\ga=1$ (and consequently $\eta=0$), we would like to stress that in the case of ghost-free infinite derivative gravity $\ph_{_{SEP}}=0$, as it may have been expected.

\subsection{Non-local gravity with non-analytic form-factors}

For the last case, we consider two models of non-local infrared extension of Einstein's GR, where form-factors are non-analytic functions of $\Box$. These theories are inspired by quantum corrections to the effective action of quantum gravity~\cite{Bravinsky,Deser:2007jk,Deser:2013uya,Belgacem:2017cqo,Tan:2018bfp,Conroy:2014eja,Woodard:2018gfj}.
\paragraph{First model:}
The first model is described by the following choice of the form-factors:
	\begin{equation}
	\mathcal{F}_1=\frac{\alpha}{\Box}\,,\,\,\,\,\mathcal{F}_2=0\,\,\,\Longrightarrow\,\,f=1\,,\,\,\,\,g=1-2\alpha\,.\label{non-local-choice1}
	\end{equation}
	The two metric potentials are infrared modifications of the Newtonian one
	\begin{eqnarray}\label{pono1}
	\phi(r)&=& -\frac{Gm}{r}\left(\frac{4\alpha-1}{3\alpha-1}\right),\\[2mm]
	\psi(r)&=&\displaystyle -\frac{Gm}{r}\left(\frac{2\alpha-1}{3\alpha-1}\right).
	\label{pono2}
	\end{eqnarray}
Since we expect $\al$ to be small, we can deduce that the Eddington-Robertson-Schiff parameter for this model is represented by
\be\label{g5}
\ga=\frac{2\al-1}{4\al-1}\simeq 1+2\al\,.
\ee
Starting from (\ref{pono1}) and (\ref{pono2}), we obtain
\be\label{q5}
\phi_{_{Q}}(r)=-\frac{\al}{3\al-1}\frac{Gm}{r}\,,
\ee
and consequently
\be\label{q55}
\varphi_{_{Q}}=\frac{\al}{3\al-1}\varphi_{_{GR}}.
\ee
The SEP violating phase takes the form
\be\label{sep5}
\varphi_{_{SEP}}=-\frac{\De m^2 G m}{E_\ell}\frac{\al}{3\al-1}\ln\lf(\frac{r_B}{r_A}\ri).
\ee

\paragraph{Second model:}
The non-local form factors for the second model are 
	\begin{equation}
	\mathcal{F}_1=\frac{\beta}{\Box^2}\,,\,\,\,\,\mathcal{F}_2=0\,\,\,\Longrightarrow\,\,f=1\,,\,\,\,\,g=1-\frac{2\beta}{\Box}\,.\label{non-local-choice2}
	\end{equation}
	In this framework, the infrared modification is not a constant, but the metric potentials show a Yukawa-like behavior
	\begin{eqnarray}
	\phi(r)&=&-\frac{4}{3}\frac{Gm}{r}\left(1-\frac{1}{4}e^{-\sqrt{3\beta}r}\right),\\[2mm]
	\psi(r)&=&-\frac{2}{3}\frac{Gm}{r}\left(1+\frac{1}{2}e^{-\sqrt{3\beta}r}\right).
	\end{eqnarray}
Also for the current non-local model, GR is recovered in the limit $\beta\to 0$. Therefore, an expansion around this parameter allows us to cast $\ga$ of Eq.~(\ref{ppn}) in the following form:
\be\label{g6}
\ga=\frac{1+\frac{1}{2}e^{-\sqrt{3\beta}r}}{2-\frac{1}{2}e^{-\sqrt{3\beta}r}}\simeq 1-\frac{2}{3}\sqrt{3\beta}r\,.
\ee
The gravitational potential associated to the purely quadratic part of this model reads
\be\label{q6}
\phi_{_{Q}}(r)=-\frac{1}{3}\frac{Gm}{r}\lf(1-e^{-\sqrt{3\beta}r}\ri)\,.
\ee
The phase related to the previous potential is given by
\bea\label{q66}
&&\varphi_{_{Q}}=\frac{\Delta m^2L_p}{2E_\ell}\Bigl\{\frac{Gm}{3 r_B}\lf(1-e^{-\sqrt{3\beta} r_B}\ri)\\[2mm]\non
&&\,\,\,\,\,\,\,\,\,\,\,\,\,\,-\frac{Gm}{3L_p}\ln\lf(\frac{r_B}{r_A}\ri)+\frac{Gm}{3L_p}\lf[\mathrm{Ei}\lf(-\sqrt{3\beta}r\ri)\ri]^{r_B}_{r_A}\Bigr\} \, .
\eea
The SEP violating phase now reads
\bea\label{sep6}
\hspace{-3mm}\varphi_{_{SEP}}=\frac{\Delta m^2 G m}{3 E_\ell}\Bigl\{\lf[\mathrm{Ei}\lf(-\sqrt{3\beta}r\ri)\ri]^{r_B}_{r_A}-\ln\lf(\frac{r_B}{r_A}\ri)\Bigr\} .
\eea
%
%Eq.\eqref{q66} can thus rewritten as
%\bea
%\varphi_{_{Q}}  =  \frac{\Delta m^2 L_p}{2 E_\ell}\frac{Gm}{3 r_B}\lf(1-e^{-\sqrt{3\beta} r_B}\ri)
%+2 \varphi_{_{SEP}} \, .
%\eea
%

%which returns the following constraint:
%\bea\label{b6}
%&&\hspace{-5mm}\Bigl|\frac{1-e^{-\sqrt{3\beta}\lf(L_p+r_A\ri)}}{L_p+r_A}-\frac{1}{L_p}\ln\lf(1+\frac{L_p}{r_A}\ri)\\[2mm]\non
%&&+\frac{1}{L_p}\lf[\mathrm{Ei}\lf(-\sqrt{3\beta}r\ri)\ri]^{L_p+r_A}_{r_A}\Bigr|<\frac{3}{Gm}.
%\eea

%%%%%%%%%%%%%%%%%%%%%%%%%%%%%%%%%%%%%%%%%%%%%%%%%%%%%
\section{Concluding Remarks} \label{conc}
In this work, we have studied neutrino oscillation within the framework of quadratic theories of gravity. Specifically, we have shown to what extent the quadratic part of the action~(\ref{quad-action}) contributes to the covariant phase $\varphi_{12}$ via the emergence of extra terms into the flavor oscillation probability. In light of this, we have stressed that it is always possible to split $\varphi_{12}$ into different terms, among which we have recognized the analogue of the flat phase $\varphi_{_0}$, the GR-induced phase $\varphi_{_{GR}}$ and the corrections pertaining to the quadratic sector $\varphi_{_{Q}}$. Calculations have been performed for neutrino oscillations
both in vacuum and matter, noticing that formulas
in the latter case can be obtained from the corresponding
equations in vacuum by accounting for the redefinitions~\eqref{mat} and~\eqref{mat2}.
 
Apart from their intrinsic theoretical relevance, 
it would be interesting to analyze our results
in connection with possible experimental applications. 
For instance, it has been shown that non-trivial gravitational 
contributions to the neutrino oscillation phase
might have significant effects in supernova explosions~\cite{Grossman}, 
for which also matter effects play an important r\^ole. 
In light of this and by exploiting the existing data 
on neutrino oscillations, the present study may provide 
an important step towards a deeper understanding of gravity, 
since it could help us to shed some light 
in the current zoo of theories, both
validating or ruling out them at a fundamental level.
This aspect, however, will be investigated in more detail
in a future work.
%Since one expects the terms related to the Einstein-Hilbert action in Eq.~(\ref{quad-action}) to produce the most significant effects from a phenomenological point of view (when dealing with Solar system experiments), it is licit to assume that the inequality~(\ref{asm}) holds true. As a consequence, this has allowed us to infer the constraint~(\ref{cons}), which can be potentially traduced in a bound for the free parameters contained in the quadratic sector of the gravitational action. 
 
Another crucial result we have pointed out is the possibility to identify a contribution associated to the violation of the strong equivalence principle in the expression of the oscillation phase. Indeed, for different gravitational potentials, $\phi\neq\psi$, we have observed that the Nordtvedt parameter $\eta$ does not vanish, which in turn implies SEP violation. This occurrence has been achieved by requiring all post-Newtonian terms of the examined models to be equivalent to the GR ones, except for the Eddington-Robertson-Schiff parameter $\ga$. A more rigorous treatment which includes the whole set of post-Newtonian expansion coefficients would require a full-fledged analysis that goes beyond the linearized approximation. However, the generality of the aforesaid outcome is not affected by the regime in which we have investigated such an intriguing issue. 

Finally, we have implemented the above reasoning on several quadratic theories of gravity. The purpose of this application is to draw the attention on the expressions for the neutrino oscillation phase $\ph_{_{Q}}$ related to the quadratic part of the action. Furthermore, we have explicitly written the contribution arising from the presence of SEP violation $\ph_{_{SEP}}$, which enters in $\ph_{_{Q}}$ and not in $\ph_{_{GR}}$, as expected. 
%It is worth mentioning that, among the models exhibited so far, the ghost-free infinite derivative gravity is the only model for which the strong equivalence principle holds. In all other cases, we have shown that $\ga\neq 1$, thus encountering a non-vanishing $\ph_{_{SEP}}$.

%As a final remark, we want to stress that in Sec.~\ref{Newsec} we have discussed a plausible explanation for the appearance of the phase term, which can be viewed as a phenomenological implication of SEP violation. The arguments in favor of this concept are based on the expression of the Dirac Hamiltonian~(\ref{dirh3}). In fact, the presence of a factor depending on the gradient of $\phi-\psi$ clearly suggests that there is an additional term to take care of when $\ga\neq 1$. The previous considerations may then be employed in different contexts as further pieces of evidence for an actual violation of the strong equivalence principle. Moreover, this extra term in neutrino phase could be plausibly related to a geometrical phase \` a la Berry~\cite{Berry} or \`a la Aharonov--Bohm~\cite{BA}.  Work in this direction is still under development~\cite{prep}.

\acknowledgments
We would like to thank M.~Blasone and G.~Lambiase for their useful comments. 
We are extremely grateful to the anonymous Referees for their inspiring suggestions, which improved the quality of the manuscript. We are indebted to Modestino for his inputs.
G.~G.~L. and L.~P. would like to thank P.~Brosio for enlightening discussions. 

%%%%%%%%%%%%%%%%%%%%%%%%%%%%%%%%%%%%%%%%%%%%%%%%%%%%%%%%%%%%%%%%%%%%%%%%%%%%%%%%%%%%%%%%

%%%%%%%%%%%%%%%%%%%%%%%%%%%%%%%%%%%%%%%%%%%%%%%%%%%%%%%%%%%%%%%%%%%%%%%%%%%%%%%%%%%%%%%%
\end{document}